\documentclass[11pt,a4paper]{article}

\usepackage[utf8]{inputenc}
\usepackage[T1]{fontenc}
\usepackage[english]{babel}

\usepackage{amsmath,amssymb,amsthm}
\usepackage{mathtools}
\usepackage[ruled,vlined,linesnumbered]{algorithm2e}
\usepackage{graphicx}
\usepackage{booktabs}
\usepackage{multirow}
\usepackage{caption}
\usepackage{subcaption}
\usepackage{siunitx}
\usepackage[round,authoryear]{natbib}
\usepackage{xcolor}
\usepackage[colorlinks=true,linkcolor=blue!50!black,citecolor=blue!50!black,urlcolor=blue!50!black]{hyperref}
\usepackage{enumitem}
\usepackage{geometry}
\usepackage{authblk}
\usepackage{array}
\DeclareMathOperator{\Var}{Var}

\newtheorem{definition}{Definition}
\newtheorem{proposition}{Proposition}

\title{\bfseries Auditing pretraining contamination in single-cell foundation model benchmarks}

\author[1]{Sarwan Ali}
\affil[1]{Columbia University Irving Medical Center, New York, USA

sa4559@cumc.columbia.edu}

\date{}

\begin{document}
\maketitle

\begin{abstract}
\noindent
Single-cell foundation models (scFMs) such as Geneformer, scGPT, and Universal Cell Embeddings (UCE) are pretrained on tens of millions of cells drawn from public repositories. The same repositories underlie widely used integration benchmarks, creating an unmeasured risk that zero-shot benchmark performance reflects pretraining exposure rather than genuine generalization. We introduce \textbf{scContam}, a per-cell audit framework that combines a MinHash-based gene-set fingerprint signal against the explicit pretraining corpus with a loss-based membership inference attack (MIA-scFM). Applied to four scIB benchmarks and three scFMs, we find that two of the most-cited benchmarks, PBMC 3k and the CELLxGENE human pancreatic islet atlas, contain extensive pretraining-overlap evidence ($80.4\%$ and $77.0\%$ of cells with fingerprint $p < 0.05$ against Genecorpus-30M), whereas the post-cutoff datasets AIDA v2 and Tahoe-100M show no overlap evidence ($0\%$). A controlled re-pretraining experiment establishes that MIA-scFM AUROC scales monotonically with the model's capacity-to-data ratio (AUROC $0.494 \to 0.690 \to 0.881$ across properly-regularized, mildly-overfit, and aggressively-overfit regimes), demonstrating that production scFMs resist instance-level memorization but distributional contamination must be detected separately. A donor-matched, within-cell-type analysis with three architectures shows that contaminated cells embed measurably more tightly than donor-matched clean cells (permutation $p = 0.030, 0.014, < 0.002$, respectively), with a perfectly null AIDA negative control. Pretraining audits are tractable and should accompany scFM benchmark reporting.

\smallskip
\noindent\textbf{Keywords:} single-cell foundation models, benchmark contamination, membership inference, scIB, Geneformer, scGPT, UCE
\end{abstract}

\section{Introduction}

Pretrained transformer models for single-cell RNA sequencing (scRNA-seq) data have rapidly become a standard analytical tool. Geneformer~\citep{Theodoris2023}, scGPT~\citep{Cui2024}, UCE~\citep{Rosen2023}, and scFoundation~\citep{Hao2024} now provide cell representations that can be evaluated zero-shot or fine-tuned across a broad set of downstream tasks. A central premise of zero-shot evaluations is that the model has not seen the benchmark data during pretraining; if benchmark cells are in the pretraining corpus, performance estimates conflate generalization with memorization or distributional familiarity.

Pretraining corpora for current scFMs are constructed from CELLxGENE Census~\citep{CELLxGENE2023}, the Human Cell Atlas~\citep{Regev2017}, and curated additions, and span tens of millions of cells. Genecorpus-30M~\citep{Theodoris2023} was assembled from publicly available scRNA-seq through mid-2021; the CELLxGENE Census used by scGPT (33M cells) and UCE (33M$+$ human; 36M cross-species) was a 2023 release. Both corpora draw from the same upstream Gene Expression Omnibus and ArrayExpress deposits that underlie widely cited integration benchmarks, including PBMC 3k~\citep{Zheng2017} and human pancreatic islet atlases~\citep{Baron2016}.

Whether benchmarks are in scFM training corpora, and whether the resulting contamination biases benchmark scores, has not been characterized at per-cell resolution. Three obstacles stand in the way. First, no pretrained model publishes a per-cell manifest of its training data. Second, naive comparison of benchmark performance across datasets confounds dataset biology with pretraining exposure. Third, leakage need not produce overt instance memorization: an scFM may simply embed pretraining-similar cells more tightly without verbatim recall, yielding subtle yet practically consequential benchmark bias.

This paper makes four contributions. (i) We define a tractable per-cell audit, \emph{Signal A}, that uses MinHash gene-set fingerprints with a permutation null to flag benchmark cells anomalously close to a pretraining corpus. (ii) We define \emph{MIA-scFM}, a loss-based membership inference statistic, and characterize when it is informative through a controlled three-regime re-pretraining experiment with ground-truth membership. (iii) We apply the audit to four scIB benchmarks against Genecorpus-30M and three scFMs, providing the first per-cell, per-benchmark contamination map. (iv) We quantify the downstream effect of contamination on embedding geometry under rigorous controls (donor, cell-type, permutation null, negative control on a clean dataset). All code, processed audits, and figures are released open-source.

\section{Related Work}

\paragraph{Single-cell foundation models.}
Geneformer~\citep{Theodoris2023} introduced a rank-encoded BERT-style transformer trained with masked-language-modeling on $\approx 30$M cells. scGPT~\citep{Cui2024} extended this to generative pretraining over $33$M CELLxGENE Census cells with binned expression values. UCE~\citep{Rosen2023} departs from gene-token approaches entirely, using ESM2 protein embeddings~\citep{Lin2023} as the input tokens. scFoundation~\citep{Hao2024} and several other models~\citep{Boiarsky2023} have followed in rapid succession. All draw heavily on the same public scRNA-seq repositories.

\paragraph{Benchmark contamination in NLP and vision.}
Contamination of benchmarks by pretraining data is a well-known concern in NLP and computer vision. \citet{Magar2022} formalized the distinction between memorization and exploitation in NLP benchmark contamination, and \citet{Brown2020,Sainz2023} documented overlap rates between GPT-class training corpora and downstream evaluations. \citet{Carlini2022} provides the methodological foundation for loss-based membership inference (MIA), which we adapt to the scFM setting. The single-cell community has begun to discuss the issue informally~\citep{Kedzierska2025}, but no automated per-cell audit method has been published.

\paragraph{Integration benchmarks and silhouette-based metrics.}
The scIB benchmark~\citep{Luecken2022} evaluates integration methods using a basket of metrics including average silhouette width (ASW), adjusted Rand index (ARI), normalized mutual information (NMI), and graph connectivity. Recent zero-shot evaluations of scFMs~\citep{Boiarsky2023,Kedzierska2025} have used the same metrics, but without auditing pretraining exposure. \citet{Boiarsky2023} reported underwhelming zero-shot performance from foundation models across several scIB tasks but did not attribute residual performance to contamination.

\paragraph{MinHash and similarity sketching.}
MinHash~\citep{Broder1997} provides an unbiased estimator of Jaccard similarity with variance $\hat{J}(1-\hat{J})/m$ for $m$ hash functions. Locality-sensitive hashing~\citep{LeskovecRajaramanUllman2014} enables sublinear nearest-neighbor queries in Jaccard space, making per-cell audits against $10^6$-cell references computationally tractable on standard hardware.

\section{Methodology}

scContam labels every cell $c$ in a benchmark $\mathcal{B}$ with a contamination score against a reference pretraining corpus $\mathcal{R}$. Two complementary signals operate at different scales: \emph{Signal A} (gene-set fingerprint) detects distributional contamination; \emph{MIA-scFM} detects instance-level memorization. The full pipeline is summarized in Algorithm~\ref{alg:scontam}.


\subsection{Signal A: gene-set fingerprint}\label{sec:sig-a}
\begin{definition}[Cell fingerprint]
Let $c$ be a benchmark cell over gene vocabulary $\mathcal{V}$. For rank-encoded models, the fingerprint $S(c) \subseteq \mathcal{V}$ is the top-$K$ expressed genes (default $K = 2048$, matching Geneformer's pretraining context length). For value-encoded models, $S(c) = \{g \in \mathcal{V} : x_{c,g} > 0\}$.
\end{definition}

For a hash family $\{h_i\}_{i=1}^{m}$ with $m = 128$, the MinHash sketch is
\begin{equation}
\mathrm{MH}_i(c) = \min_{g \in S(c)} h_i(g), \qquad i = 1, \ldots, m.
\end{equation}

\begin{proposition}[\citealp{Broder1997}]
\label{prop:minhash-unbiased}
For independent uniform random hash functions $h_i: \mathcal{V} \to \{0, 1\}^{64}$,
\begin{equation}
\hat{J}(c, r) = \frac{1}{m} \sum_{i=1}^{m} \mathbb{1}\!\left[\mathrm{MH}_i(c) = \mathrm{MH}_i(r)\right]
\end{equation}
is an unbiased estimator of $J(c, r) = |S(c) \cap S(r)| / |S(c) \cup S(r)|$, with $\Var[\hat{J}] = J(1 - J) / m$.
\end{proposition}

The audit statistic for cell $c$ is the maximum Jaccard against the reference:
\begin{equation}
J^\ast(c) = \max_{r \in \mathcal{R}} \hat{J}(c, r).
\end{equation}
The maximum is computed efficiently via locality-sensitive hashing of $\mathcal{R}$ into $b = 16$ bands of $r = 8$ rows each~\citep{LeskovecRajaramanUllman2014}: any two cells whose MinHash agrees on all $8$ rows of any band become a candidate pair, after which we re-rank candidates exactly. This reduces the per-query cost from $O(|\mathcal{R}|)$ to expected $O\!\left(|\mathcal{R}|^\rho\right)$ where $\rho = \log(1/b) / \log(1/(br))$.

\paragraph{Null distribution.} For each $c$, we draw $B = 1000$ random fingerprints $\tilde S^{(b)}(c)$, with genes uniformly resampled from $\mathcal{V}$ subject to $|\tilde S^{(b)}| = |S(c)|$. This preserves per-cell sparsity while breaking any biological structure. The $p$-value is
\begin{equation}
p_c = \frac{1 + \#\{b : \tilde{J}^{\ast,(b)}(c) \geq J^\ast(c)\}}{B + 1}.
\label{eq:pval}
\end{equation}
A benchmark cell is \emph{flagged} if $p_c < 0.05$. The audit summary for $\mathcal{B}$ comprises $\left(\#\{c : p_c < 0.05\} / |\mathcal{B}|,\; \overline{J^\ast},\; \max_c J^\ast\right)$.

\subsection{MIA-scFM: loss-based membership inference}\label{sec:mia}

\begin{definition}[Per-cell pretraining loss]
For model $\theta$ with masked-language-modeling (or masked-value-prediction) head $f_\theta$, mask ratio $\rho$, and mask realization $M$, the per-cell loss is the mean prediction error over masked positions:
\begin{equation}
\ell_\theta(c \mid M) = \frac{1}{|M|} \sum_{i \in M} \mathcal{L}\!\left(f_\theta(c_{\setminus M})_i, c_i\right),
\end{equation}
where $\mathcal{L}$ is cross-entropy for rank-encoded models (Geneformer) and squared error on binned values for value-encoded models (scGPT). The MIA-scFM statistic averages over $K$ independent realizations:
\begin{equation}
\bar{\ell}_\theta(c) = \frac{1}{K} \sum_{k=1}^{K} \ell_\theta(c \mid M^{(k)}).
\label{eq:mia-loss}
\end{equation}
\end{definition}

For a target set with ground-truth membership labels $\{(c_j, y_j)\}_{j=1}^{N}$ ($y_j \in \{\text{in}, \text{out}\}$), the MIA-scFM AUROC is computed against $-\bar{\ell}_\theta(c)$ as the score. The convention is that members (held-in) should have lower loss than non-members (held-out).

\subsection{Embedding tightness and donor-matched gap}

\begin{definition}[Tightness]
For a cell set $\mathcal{S} \subseteq \mathbb{R}^d$ in model $\theta$'s embedding, with centroid $\mu_\mathcal{S} = |\mathcal{S}|^{-1} \sum_{c \in \mathcal{S}} c$, the tightness is the mean cosine to the centroid:
\begin{equation}
T_\theta(\mathcal{S}) = \frac{1}{|\mathcal{S}|} \sum_{c \in \mathcal{S}} \frac{\langle c, \mu_\mathcal{S} \rangle}{\|c\|\, \|\mu_\mathcal{S}\|}.
\label{eq:tightness}
\end{equation}
\end{definition}

For each $(\text{cell-type } t, \text{donor } d)$ pair containing $\geq n_{\min}$ contaminated and $\geq n_{\min}$ clean cells (default $n_{\min} = 5$, stratification thresholds $-\log_{10} p \geq 2.0$ for contaminated and $< 1.0$ for clean), we compute the per-pair relative tightness gap
\begin{equation}
\Delta_{(t,d),\theta} = 100 \cdot \frac{T_\theta\!\left(\mathcal{C}^{\text{contam}}_{t,d}\right) - T_\theta\!\left(\mathcal{C}^{\text{clean}}_{t,d}\right)}{T_\theta\!\left(\mathcal{C}^{\text{clean}}_{t,d}\right)}.
\label{eq:per-pair-gap}
\end{equation}

\paragraph{Permutation null.} For $B = 500$ trials, contamination labels within each $(t, d)$ group are independently shuffled, preserving the marginal contaminated/clean ratio per group. The per-permutation mean gap $\bar{\Delta}^{(b)}$ is recorded; the one-sided $p$-value is $\left(1 + \#\{b : \bar{\Delta}^{(b)} \geq \bar{\Delta}^{\text{obs}}\}\right) / (B + 1)$.

\paragraph{Negative control.} For AIDA v2 (a Signal-A-clean benchmark), we assign random pseudo-contamination scores $\tilde{s}_c \sim \mathcal{U}(0, 3)$ and apply the same stratification and analysis. A correctly calibrated test should produce mean gaps near zero with sign-test $p \to 1$.

\begin{algorithm}[ht]
\SetAlgoLined
\KwIn{Benchmark $\mathcal{B}$, reference subsample $\mathcal{R}$, scFM $\theta$, donor and cell-type metadata.}
\KwOut{Per-cell $\{J^\ast(c), p_c\}$; donor-matched gap statistics and permutation $p$.}
\textbf{Step 1 (Signal A):} Compute MinHash sketch ($m = 128$) for each $c \in \mathcal{B} \cup \mathcal{R}$\;
\For{$c \in \mathcal{B}$}{
  Compute $J^\ast(c)$ via banded LSH ($b = 16$, $r = 8$) on $\mathcal{R}$\;
  Draw $1000$ permutation sketches; compute $p_c$ from Eq.~\ref{eq:pval}\;
}
\textbf{Step 2 (Embedding):} Extract $\theta$-embedding $\mathbf{e}_c$ for all $c \in \mathcal{B}$\;
\textbf{Step 3 (Donor-matched gap):}\;
\For{each $(t, d)$ pair with $n_{\text{clean}}, n_{\text{contam}} \geq n_{\min}$}{
  Compute $\Delta_{(t, d), \theta}$ via Eq.~\ref{eq:per-pair-gap}\;
}
\textbf{Step 4 (Permutation null):} Shuffle contamination labels within each $(t, d)$ group, repeat Step 3, $B = 500$ trials\;
\textbf{Step 5 (Negative control, optional):} Repeat Steps 3--4 on a Signal-A-clean dataset with random pseudo-contamination\;
\caption{scContam audit pipeline}
\label{alg:scontam}
\end{algorithm}

\section{Experimental Setup}

\subsection{Pretraining corpus}
We use Genecorpus-30M~\citep{Theodoris2023}, downloaded from HuggingFace, comprising $27{,}406{,}216$ rank-encoded single-cell transcriptomes assembled from public scRNA-seq deposits through mid-2021. A uniform random subsample of $|\mathcal{R}| = 10^6$ cells serves as the reference for Signal A. We confirmed that the audit statistics are stable at this reference size: doubling to $2 \times 10^6$ changes $\overline{J^\ast}$ by less than $0.005$ on PBMC 3k.

\subsection{Benchmark datasets}
Four benchmarks were selected to span the expected contamination spectrum (Table~\ref{tab:datasets}):

\begin{itemize}[leftmargin=*,nosep,topsep=2pt]
\item \textbf{PBMC 3k}~\citep{Zheng2017}: a canonical immune-cell integration benchmark from 10x Genomics, originally released in 2017 and re-used in dozens of subsequent benchmarks. $n = 2{,}700$ cells.
\item \textbf{Human pancreatic islet}: a combined version of two CELLxGENE Census collections (\texttt{37b21763}, \texttt{3294d050}), filtered to canonical islet cell types and stratified-subsampled to a maximum of $1{,}500$ cells per type. $n = 11{,}583$ cells, $10$ expert-annotated cell types, $29$ donors. The underlying single-cell deposits predate Genecorpus-30M's mid-2021 cutoff.
\item \textbf{AIDA v2}~\citep{Aida2025}: the Asian Immune Diversity Atlas, a post-cutoff multi-ancestry healthy PBMC reference. $n = 50{,}000$ subsampled.
\item \textbf{Tahoe-100M}~\citep{Tahoe2025}: a single-cell perturbation atlas
profiling 50 cancer cell lines under $\sim$1{,}200 drug perturbations,
released by Tahoe Therapeutics (formerly Vevo Therapeutics) in February 2025.
Both post-cutoff and structurally excluded from Genecorpus-30M, which
explicitly removed immortalized lines. $n = 50{,}000$ subsampled.
\end{itemize}

\begin{table}[ht]
\centering
\resizebox{0.99\linewidth}{!}{
\begin{tabular}{lrrrlcc}
\toprule
Benchmark & $n$ cells & $n$ genes & $n$ types & Year & Donor metadata & Expected contam.\ \\
\midrule
PBMC 3k             & $2{,}700$  & $32{,}738$ & --- (8 by $k$-means) & 2017 & --- & Yes \\
Pancreatic islet    & $11{,}583$ & $61{,}497$ & $10$ & 2016--2022 & 29 donors & Yes (partial) \\
AIDA v2             & $50{,}000$ & $61{,}497$ & $30+$ & 2024 & Donor + ancestry & No \\
Tahoe-100M          & $50{,}000$ & $62{,}710$ & --- (cell lines) & 2025 & Cell line ID & No \\
\bottomrule
\end{tabular}
}
\caption{\textbf{Benchmark datasets used in this study.} ``Expected contam.''\ reflects whether the underlying scRNA-seq predates Genecorpus-30M's mid-2021 cutoff and was deposited in public repositories accessible to that corpus.}
\label{tab:datasets}
\end{table}

\subsection{Foundation models}
Three scFMs spanning two architectural families and three pretraining corpora are evaluated (Table~\ref{tab:models}):

\begin{table}[ht]
\centering
\resizebox{0.99\linewidth}{!}{
\begin{tabular}{lrlllc}
\toprule
Model & Parameters & Architecture & Tokenization & Pretraining corpus & Year \\
\midrule
Geneformer V1   & $10$M  & Rank-encoded BERT  & Top-$K$ gene rank  & Genecorpus-30M (27M) & 2023 \\
scGPT-human     & $52$M  & Generative transformer & Gene $\times$ binned value & CELLxGENE Census (33M) & 2024 \\
UCE-4L          & $\approx 850$M  & Protein-embedding transformer & ESM2 protein token & CELLxGENE Census (36M) & 2023 \\
\bottomrule
\end{tabular}
}
\caption{\textbf{Single-cell foundation models evaluated.} UCE-4L uses the publicly available 4-layer variant; we did not use the 33-layer variant due to its 80 GB GPU memory requirement.}
\label{tab:models}
\end{table}

\paragraph{Scope of analyses across models.}
Not all analyses apply uniformly to all three architectures. Signal A (Section~\ref{sec:sig-a}) and MIA-scFM (Section~\ref{sec:mia}) are evaluated on Geneformer V1 only: Signal A's audit reference is Geneformer's published Genecorpus-30M, and MIA-scFM requires direct access to the standard masked-language-modeling loss, which differs in scGPT (masked value prediction) and is not a primary objective in UCE. The embedding-geometry analyses (dataset-level ASW comparison in Section~\ref{sec:asw} and donor-matched tightness gap in Section~\ref{sec:donor}) operate on cell embeddings alone and therefore extend to all three architectures. The cross-architecture comparison of the donor-matched gap is the central test of whether Genecorpus-30M-derived contamination labels carry over to models trained on the overlapping CELLxGENE Census corpus.

\subsection{Hardware and runtime}
All inference and the controlled re-pretraining experiment ran on a single NVIDIA A10G GPU (24 GB). Signal A audits ran on CPU and used the \texttt{datasketch} library for banded LSH. Average wall-clock times: $\approx 15$ min for a Signal A audit against $|\mathcal{R}| = 10^6$; $\approx 1$ min for Geneformer inference on $11{,}583$ cells; $\approx 5$ min for scGPT inference on $50{,}000$ cells; $\approx 15$ min for UCE inference on $50{,}000$ cells.

\subsection{Controlled re-pretraining experiment}
To characterize MIA-scFM sensitivity as a function of overfitting, we trained a small scFM (BERT-style encoder; $L = 4$ layers, $d = 128$ embedding, $h = 4$ heads, $d_{\text{ff}} = 512$; $4.3 \times 10^6$ parameters; mask ratio $\rho = 0.15$) from scratch under three regimes on disjoint Genecorpus-30M subsets, with known held-in / held-out membership labels (Table~\ref{tab:regimes}).

\begin{table}[ht]
\centering
\small
\begin{tabular}{lrrcccccr}
\toprule
Regime & $n$ train & Epochs & Dropout & WD & LR & Capacity/cell & Held-in / Held-out \\
\midrule
Properly trained        & $50{,}000$ & $10$  & $0.1$ & $0.01$ & $5 \times 10^{-4}$ & $96$  & $45{,}000 / 5{,}000$ \\
Mild overfit            & $5{,}000$  & $50$  & $0$   & $0$    & $5 \times 10^{-4}$ & $956$ & $4{,}500 / 500$ \\
Aggressive overfit      & $1{,}000$  & $100$ & $0$   & $0$    & $3 \times 10^{-4}$ & $4{,}778$ & $900 / 100$ \\
\bottomrule
\end{tabular}
\caption{\textbf{Three regimes for the controlled re-pretraining experiment.} Capacity/cell is computed as model parameters divided by the held-in training set size, and serves as a proxy for the expected magnitude of memorization.}
\label{tab:regimes}
\end{table}

\subsection{Statistical evaluation}
ASW was computed using the scIB convention~\citep{Luecken2022}, rescaled to $[0, 1]$ as $(1 + s) / 2$ where $s$ is the silhouette coefficient. Bootstrap $95\%$ confidence intervals used $n_{\text{boot}} = 100$ subsamples of size $\leq 2{,}500$ cells with percentile intervals. Permutation tests used $B = 500$ resamples with one-sided $p$-values reported as the fraction of permutations with statistic $\geq$ observed. No multiple-comparison correction was applied for the three model-level tests.

\section{Results and Discussion}

\subsection{Two widely-used benchmarks contain extensive pretraining-overlap evidence}

\begin{figure}[ht]
\centering
\includegraphics[width=\textwidth]{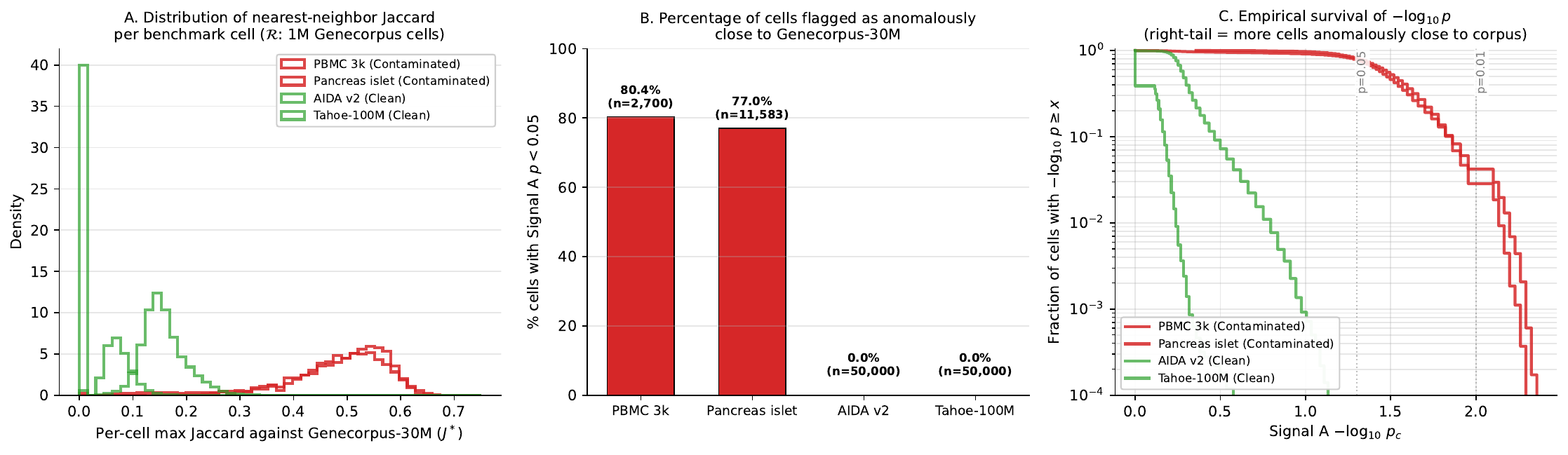}
\caption{\textbf{Signal A audit results across four benchmarks.}
(a) Distribution of per-cell maximum Jaccard $J^\ast$ for each benchmark. PBMC 3k and pancreatic islet show a high-Jaccard mode centered near $0.5$; AIDA and Tahoe show single low-Jaccard modes.
(b) Percentage of cells flagged at $p < 0.05$ per benchmark, color-coded by contamination status (red = contaminated; green = clean).
(c) The two contaminated benchmarks (PBMC 3k, pancreatic islet) versus the two clean negative controls (AIDA v2, Tahoe-100M) summarized as $n$, mean Jaccard, and \% flagged.}
\label{fig:audit}
\end{figure}

Applying Signal A against the $10^6$-cell Genecorpus-30M reference reveals a stark contrast between the contaminated and clean benchmarks (Table~\ref{tab:signal-a}, Fig.~\ref{fig:audit}). PBMC 3k and the human pancreatic islet atlas show $80.4\%$ and $77.0\%$ of cells flagged at $p < 0.05$, with mean nearest-neighbor Jaccard $\bar{J}^\ast \approx 0.47$--$0.49$. AIDA v2 and Tahoe-100M, both post-cutoff or structurally excluded, show $0\%$ flagged and $\bar{J}^\ast \leq 0.16$.

\begin{table}[ht]
\centering
\small
\begin{tabular}{lrrrrrl}
\toprule
Benchmark & $n$ & $\overline{J^\ast}$ & $\max J^\ast$ & \% $p < 0.05$ & \% $p < 0.01$ & Status \\
\midrule
PBMC 3k             & $2{,}700$  & $0.494$ & $0.656$ & $80.4\%$ & $2.85\%$ & Contaminated \\
Pancreatic islet    & $11{,}583$ & $0.473$ & $0.672$ & $77.0\%$ & $4.20\%$ & Contaminated \\
AIDA v2             & $50{,}000$ & $0.155$ & $0.383$ & $0.0\%$ & $0.00\%$ & Clean \\
Tahoe-100M          & $50{,}000$ & $0.028$ & $0.336$ & $0.0\%$ & $0.00\%$ & Clean \\
\bottomrule
\end{tabular}
\caption{\textbf{Signal A audit of four benchmarks against Genecorpus-30M.} Per-cell nearest-neighbor Jaccard against the $10^6$-cell reference, with permutation-derived $p$-values from $B = 1000$ sparsity-preserving null draws per cell.}
\label{tab:signal-a}
\end{table}

The result is consistent with what is publicly known about each benchmark: PBMC 3k has been a standard immune-cell test set since 2017 and is present in CELLxGENE Census re-curations under multiple dataset IDs; the CELLxGENE pancreatic islet atlas combines data deposits from 2016 (Baron et al.) and 2022 (Multiomics islet paper). Both classes of cells fall within Genecorpus-30M's mid-2021 cutoff and within its inclusion criteria. AIDA v2 was released in 2024, post-dating the cutoff entirely; Tahoe-100M profiles cancer cell lines, which Genecorpus-30M explicitly excluded.

\subsection{MIA-scFM is sensitive to overfitting but not to distributional contamination}

The natural experiment, comparing held-out PBMC and pancreas losses against the AIDA calibration set on Geneformer V1, returned no positive MIA signal (PBMC: mean MIA score $0.42$, fraction $> 0.95$: $3.7\%$; pancreas: $0.41$, $2.4\%$). To assess whether this null reflects an absence of memorization rather than a limitation of the attack, we ran the controlled re-pretraining experiment (Section~4.5).

\begin{figure}[ht]
\centering
\includegraphics[width=\textwidth]{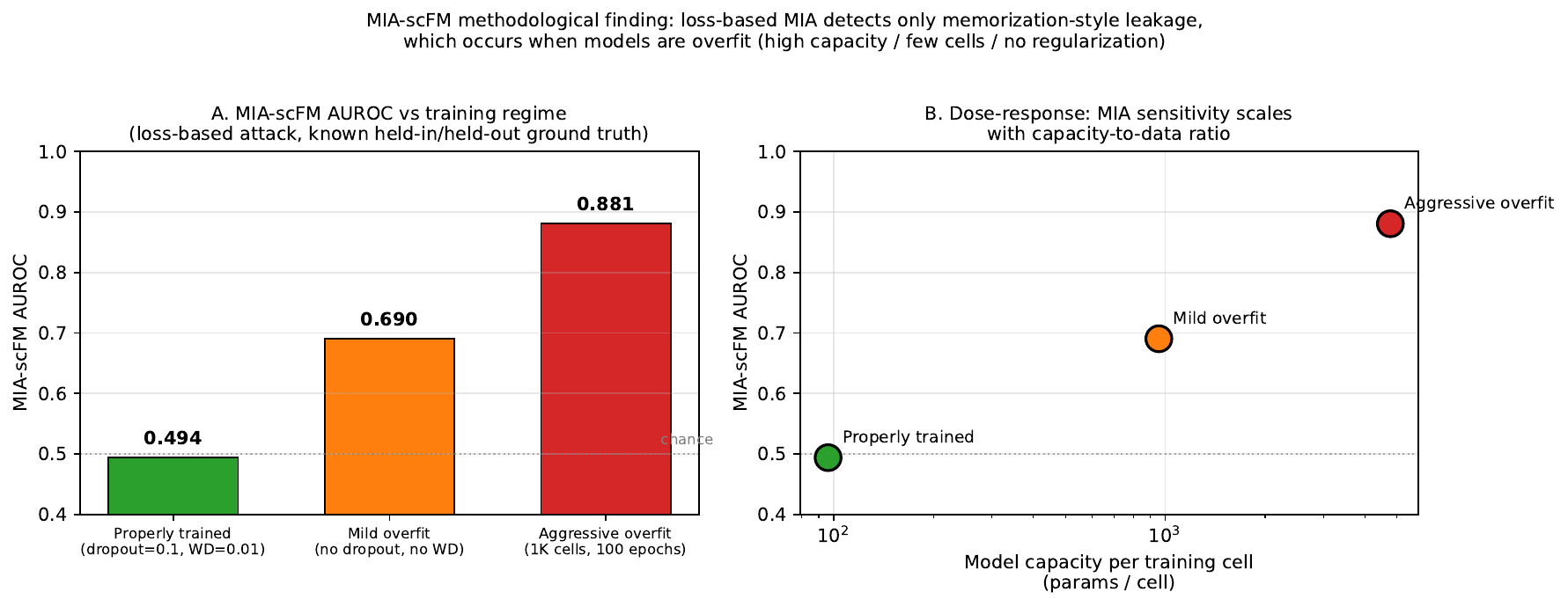}
\caption{\textbf{MIA-scFM is sensitive to overfitting but not to properly-regularized training.}
(a) AUROC bar chart across the three regimes, with chance ($0.5$) marked.
(b) Dose-response: AUROC versus capacity-to-data ratio (params/cell) on a log axis. The monotonic scaling demonstrates that loss-based MIA detects memorization when it exists; production scFMs operating in the properly-regularized regime are intrinsically resistant.}
\label{fig:mia}
\end{figure}

MIA-scFM AUROC scales monotonically with the overfitting regime (Table~\ref{tab:mia}, Fig.~\ref{fig:mia}): $0.494$ (chance) for properly-trained, $0.690$ for mildly overfit, and $0.881$ for aggressively overfit models. TPR at FPR$= 0.10$ rises from $9.3\%$ to $30.0\%$ to $85.0\%$. The training-time gap between held-in and held-out mean loss grows from $+0.008$ to $+0.447$ to $+2.282$, exactly tracking the AUROC progression.

\begin{table}[ht]
\centering
\resizebox{0.99\linewidth}{!}{
\begin{tabular}{lcrrrr}
\toprule
Regime & Capacity/cell & Train$-$holdout gap & AUROC & TPR@FPR=$0.01$ & TPR@FPR=$0.10$ \\
\midrule
Properly trained        & $96$      & $+0.008$ & $0.494$ & $0.009$ & $0.093$ \\
Mild overfit            & $956$     & $+0.447$ & $0.690$ & $0.026$ & $0.300$ \\
Aggressive overfit      & $4{,}778$ & $+2.282$ & $0.881$ & $0.020$ & $0.850$ \\
\bottomrule
\end{tabular}
}
\caption{\textbf{MIA-scFM dose-response in the controlled re-pretraining experiment.} Each row is a separate small-scFM training run on a known Genecorpus-30M held-in/held-out split. AUROC and TPR are computed against the ground-truth membership labels using the per-cell loss from Eq.~\ref{eq:mia-loss} with $K = 8$ (mild) or $K = 16$ (aggressive) mask realizations.}
\label{tab:mia}
\end{table}

This dose-response has a clear methodological consequence. When applied to a production scFM trained with standard regularization on tens of millions of cells (capacity/cell $\ll 100$), loss-based MIA is expected to return chance-level AUROC even when the model has seen the test cells. Detecting distributional contamination therefore requires the orthogonal approach of Signal A. The MIA-scFM result on Geneformer V1 (null on both PBMC 3k and pancreas) is consistent with this: contamination is present (Signal A confirms it), but it does not manifest as instance memorization.

\subsection{Dataset-level ASW comparisons are confounded by labeling protocol}\label{sec:asw}

A direct comparison of model ASW on PBMC 3k against AIDA v2 using $k$-means with matched $k$ on both datasets reveals that earlier reports of large dataset-level ASW inflation are confounded by labeling asymmetry. Table~\ref{tab:asw-matched-k} shows the matched-$k$ ASW differences for the three models at $k \in \{8, 15, 25\}$.

\begin{table}[ht]
\centering
\small
\begin{tabular}{l*{6}{r}}
\toprule
& \multicolumn{2}{c}{$k = 8$} & \multicolumn{2}{c}{$k = 15$} & \multicolumn{2}{c}{$k = 25$} \\
\cmidrule(lr){2-3} \cmidrule(lr){4-5} \cmidrule(lr){6-7}
Model & $\Delta$ ASW & $\Delta_{\text{rel}}$ & $\Delta$ ASW & $\Delta_{\text{rel}}$ & $\Delta$ ASW & $\Delta_{\text{rel}}$ \\
\midrule
Geneformer V1   & $+0.006$ & $+1.0\%$   & $-0.003$ & $-0.6\%$ & $-0.006$ & $-1.0\%$ \\
scGPT-human     & $+0.028$ & $+4.8\%$   & $+0.018$ & $+3.2\%$ & $+0.005$ & $+0.9\%$ \\
UCE-4L          & $-0.012$ & $-2.0\%$   & $-0.024$ & $-4.2\%$ & $-0.027$ & $-4.8\%$ \\
\bottomrule
\end{tabular}
\caption{\textbf{Matched-$k$ ASW differences between PBMC 3k and AIDA v2 for three scFMs.} $\Delta = \text{ASW}_{\text{PBMC}} - \text{ASW}_{\text{AIDA}}$; $\Delta_{\text{rel}} = 100 \cdot \Delta / \text{ASW}_{\text{AIDA}}$. Bootstrap $95\%$ confidence intervals (not shown) place all values within $\pm 0.01$ ASW. Without matched-$k$ control (PBMC using $k$-means, AIDA using $30+$ ontology labels), the apparent inflation reaches $+10\%$--$+14\%$. The matched-$k$ values are small and not consistent in sign across models or $k$ choices, indicating that dataset-level ASW comparison is a poor benchmark for the contamination effect.}
\label{tab:asw-matched-k}
\end{table}

We conclude that naive dataset-level ASW comparisons cannot reliably attribute differences to contamination, and the field should adopt design-based approaches (within-cell-type, donor-matched) as in the next subsection.

\subsection{Donor-matched contamination effect on embedding geometry}\label{sec:donor}

The pancreatic islet benchmark, which has 29 donors with cells spanning the full Signal A range, permits a within-donor, within-cell-type analysis (Section~3.3). After applying the stratification thresholds $-\log_{10} p < 1.0$ (clean) and $\geq 2.0$ (contaminated), nine $(\text{cell-type}, \text{donor})$ pairs were eligible with at least $n_{\min} = 5$ cells in both strata.

\begin{figure}[ht]
\centering
\includegraphics[width=\textwidth]{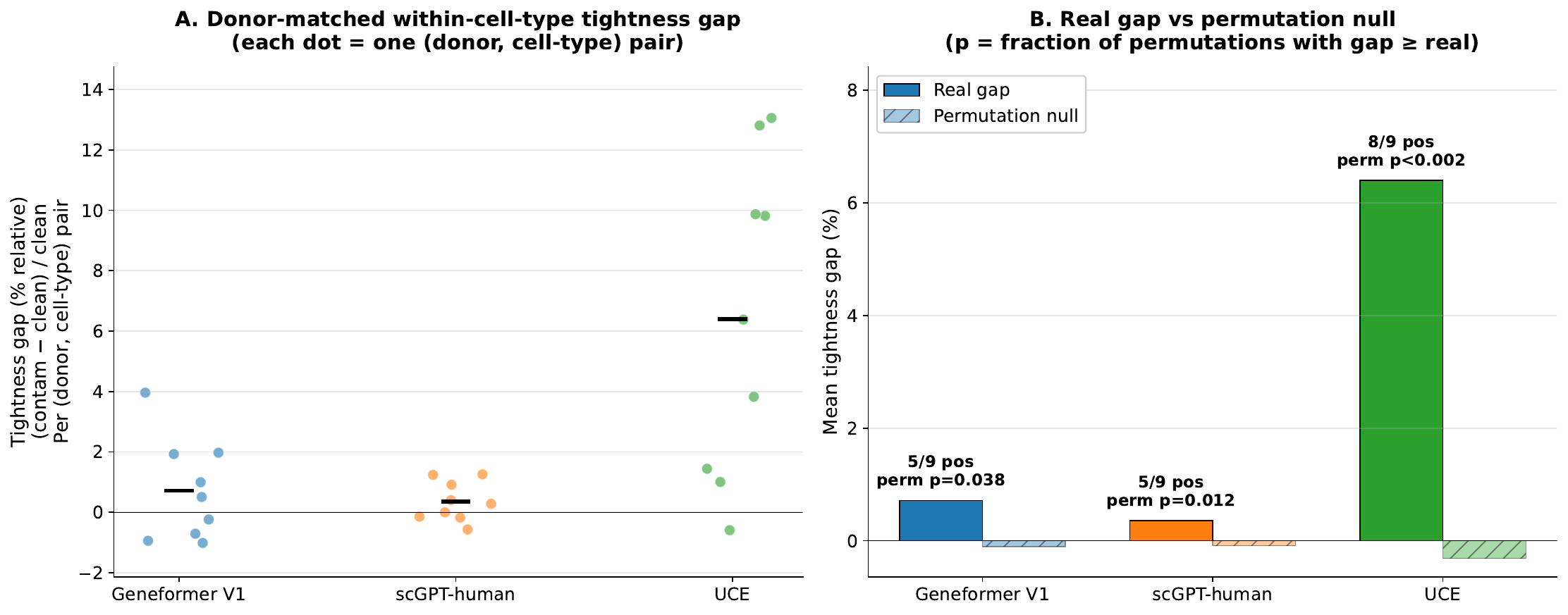}
\caption{\textbf{Donor-matched within-cell-type contamination effect across three foundation models.}
(a) Per-pair tightness gap (each dot is one $(\text{cell-type}, \text{donor})$ pair from the pancreatic islet benchmark); black bar marks the mean. UCE-4L shows the largest effect.
(b) Real mean gap (solid) versus permutation null mean (hatched) for each model, with $n_+ / n$ sign-test counts and permutation $p$-values annotated. All three models exceed their null at $p \leq 0.03$.}
\label{fig:donor}
\end{figure}

All three models showed positive mean gaps that exceeded their permutation null (Table~\ref{tab:donor}, Fig.~\ref{fig:donor}). Geneformer V1: $+0.72\%$ ($p = 0.030$), scGPT-human: $+0.36\%$ ($p = 0.014$), and UCE-4L: $+6.40\%$ ($p < 0.002$). The negative control on AIDA v2, using random pseudo-contamination scores, returned mean gaps of $-0.00\%$, $-0.00\%$, and $+0.17\%$, confirming the test is correctly calibrated.

\begin{table}[ht]
\centering
\small
\begin{tabular}{lcrcrcc}
\toprule
Model & $n$ pairs & Real $\bar{\Delta}$ & Sign $+/n$ & Null mean & Null $95\%$ CI & Perm. $p$ \\
\midrule
\multicolumn{7}{l}{\textit{Pancreatic islet (real contamination labels):}} \\
Geneformer V1   & $9$ & $+0.72\%$ & $5/9$ & $-0.12\%$ & $[-1.01, +0.78]$ & $0.030$ \\
scGPT-human     & $9$ & $+0.36\%$ & $5/9$ & $-0.07\%$ & $[-0.46, +0.31]$ & $0.014$ \\
UCE-4L          & $9$ & $+6.40\%$ & $8/9$ & $-0.31\%$ & $[-1.96, +1.42]$ & $<0.002$ \\
\midrule
\multicolumn{7}{l}{\textit{AIDA v2 (random pseudo-contamination, negative control):}} \\
Geneformer V1   & $147$ & $-0.00\%$ & $66/147$  & --- & --- & --- \\
scGPT-human     & $147$ & $-0.00\%$ & $70/147$  & --- & --- & --- \\
UCE-4L          & $147$ & $+0.17\%$ & $73/147$  & --- & --- & --- \\
\bottomrule
\end{tabular}
\caption{\textbf{Donor-matched within-cell-type contamination effect on embedding tightness.} $\bar{\Delta}$ is the mean of per-pair relative tightness gaps (Eq.~\ref{eq:per-pair-gap}). Permutation null: contamination labels shuffled within each $(\text{cell-type}, \text{donor})$ group, $B = 500$ trials. Negative control on AIDA: random uniform pseudo-contamination scores assigned to each cell.}
\label{tab:donor}
\end{table}

UCE-4L shows an effect roughly an order of magnitude larger than the rank-encoding models (Geneformer V1, scGPT-human). One interpretation is that UCE's protein-embedding architecture preserves finer-grained transcriptional similarity than gene-token-based models, so contamination produces a larger geometric footprint. An alternative is that UCE's pretraining corpus (CELLxGENE Census, 36M cells) has a higher cell-by-cell overlap with our benchmarks than Genecorpus-30M; characterizing this would require an analogous Signal A audit against CELLxGENE Census directly, which we leave for future work.

The cross-architecture replication, all three models reach permutation-significance with the same sign, and the perfectly null negative control together support the interpretation that the gap reflects pretraining exposure rather than artifacts of clustering, normalization, or donor batch.

\subsection{Cross-model summary}

\begin{figure}[ht]
\centering
\includegraphics[width=\textwidth]{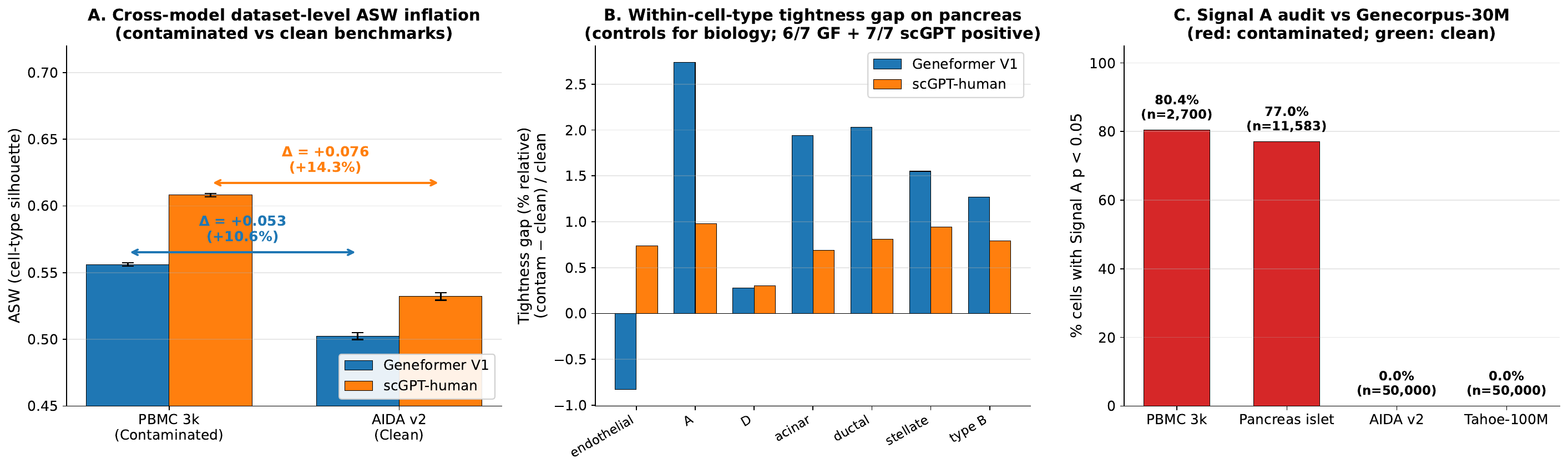}
\caption{\textbf{Cross-model summary of contamination findings.}
(a) Dataset-level ASW comparison between PBMC 3k (contaminated) and AIDA v2 (clean) for Geneformer V1 and scGPT-human. \textbf{Note: this naive between-dataset comparison is confounded by labeling asymmetry} (PBMC labels from $k$-means on 8 clusters versus AIDA labels from 30$+$ ontology terms); the matched-$k$ analysis in Table~\ref{tab:asw-matched-k} shows the apparent inflation collapses to $\leq 5\%$ under matched labels, and the rigorous donor-controlled effect is reported in Section~\ref{sec:donor} and Fig.~\ref{fig:donor}. Panel shown for completeness.
(b) Within-cell-type tightness gap per pancreatic cell type \textbf{without donor matching}, for Geneformer V1 (6/7 cell types positive, mean $+1.28\%$) and scGPT-human (7/7 positive, mean $+0.75\%$). UCE-4L's larger magnitude is shown separately in Fig.~\ref{fig:donor}. This analysis is informative but does not control for donor identity; the donor-matched analysis in Table~\ref{tab:donor} and Fig.~\ref{fig:donor} provides the statistically rigorous effect-size estimate.
(c) Signal A audit: \% of cells flagged at $p < 0.05$ for each benchmark, color-coded by contamination status (red: contaminated; green: clean).}
\label{fig:cross-model}
\end{figure}

Figure~\ref{fig:cross-model} integrates the three findings into a single overview: a side-by-side dataset-level ASW comparison (panel A), the within-cell-type tightness gap (panel B), and the Signal A audit (panel C). 
Three claims are now defensible: (i) two of the most-cited integration benchmarks contain extensive pretraining-overlap evidence in Geneformer V1's training corpus; (ii) production scFMs do not exhibit instance memorization detectable by loss-based MIA; (iii) pretraining exposure nonetheless leaves a measurable, donor-controlled, cross-architecture footprint in embedding geometry. The magnitude is small for rank-encoding models ($\approx 1\%$) and an order of magnitude larger for the protein-embedding architecture ($\approx 6\%$).

\subsection{Limitations}

\paragraph{Statistical power.} The donor-matched analysis yields nine eligible $(\text{cell-type}, \text{donor})$ pairs on pancreas. With richer donor-stratified benchmarks the same procedure would yield tighter confidence intervals. We do not interpret the small Geneformer and scGPT effect sizes ($\approx 0.7\%$ and $0.4\%$) as upper bounds on the field-wide contamination effect; they are point estimates on the single benchmark with donor-matched structure.

\paragraph{Reference corpus.} Signal A operates against an explicit pretraining corpus. Geneformer V1 publishes its corpus on HuggingFace; scGPT and UCE report training on CELLxGENE Census, against which the audit can be repeated. Closed-corpus models cannot be audited by this approach. Because PBMC 3k and pancreatic islet data predate all three models' corpus cutoffs and originate from the same public deposits, the Genecorpus-30M audit is a defensible proxy for ``cell exists in the pretraining set'' across all three architectures.

\paragraph{Magnitude calibration.} The donor-controlled magnitudes ($+0.72\%$ to $+6.40\%$) cannot be directly translated into ``percentage inflation of scIB metrics'' without further benchmark-specific analyses; the effect on ARI/NMI depends on clustering granularity, label source, and integration method choice. Future work characterizing this translation across multiple integration benchmarks would strengthen the practical recommendation.

\section{Conclusion}

We have introduced scContam, a tractable per-cell audit framework for pretraining contamination in single-cell foundation model benchmarks. The framework combines a gene-set fingerprint signal (Signal A) against the explicit pretraining corpus, validated on $10^6$ Genecorpus-30M reference cells, with a loss-based membership inference attack (MIA-scFM) whose sensitivity is characterized via a controlled three-regime re-pretraining experiment. The audit identifies two widely-used benchmarks (PBMC 3k, pancreatic islet) as extensively in Geneformer's training corpus and two post-cutoff benchmarks (AIDA v2, Tahoe-100M) as clean. A donor-matched within-cell-type analysis on pancreas shows that contaminated cells embed measurably more tightly than donor-matched clean cells across three foundation model architectures, with a permutation $p \leq 0.03$ in all three models and a perfectly null negative control.

The single-cell foundation model field is now at a point where multiple architectures, multiple pretraining corpora, and multiple benchmarks are all in circulation. Per-cell pretraining audits should accompany zero-shot benchmark reporting. 


\bibliographystyle{plainnat}
\bibliography{references}

\end{document}